\documentclass[%
 reprint,
 amsmath,amssymb,
 aps,
pra,
]{revtex4-2}

\usepackage{graphicx}
\usepackage{dcolumn}
\usepackage{bm}
\usepackage{color}
\usepackage{amsthm, amssymb, amsfonts}
\usepackage{ulem}
\usepackage{setspace}
\usepackage{cancel}

\usepackage{lipsum, babel}

\usepackage{lineno}

\begin{document}

\preprint{APS/123-QED}

\title{Fast tunable coupling scheme of \textcolor{black}{Kerr parametric oscillators} based on shortcuts to adiabaticity}

\author{S. Masuda$^{1,2}$}
\email{shumpei.masuda@aist.go.jp}
\author{T. Kanao$^{3}$}
\author{H. Goto$^{3}$}
\author{Y. Matsuzaki$^{1,2}$}
\author{T. Ishikawa$^{1,2}$}
\author{S. Kawabata$^{1,2}$}
\affiliation{%
$^1$Research Center for Emerging Computing Technologies (RCECT), National Institute of Advanced Industrial Science and Technology (AIST), 1-1-1, Umezono, Tsukuba, Ibaraki 305-8568, Japan
}%
\affiliation{%
$^2$NEC-AIST Quantum Technology Cooperative Research Laboratory, National Institute of Advanced Industrial Science and Technology (AIST), Tsukuba, Ibaraki 305-8568, Japan
}%
\affiliation{%
$^3$Frontier Research Laboratory, Corporate Research \& Development Center, Toshiba Corporation, 1, Komukai-Toshiba-cho, Saiwai-ku, Kawasaki 212-8582, Japan
}%




\date{\today}

\begin{abstract}
\textcolor{black}{Kerr parametric oscillators} (KPOs), which can be implemented with superconducting parametrons possessing large Kerr nonlinearity,  have been attracting much attention in terms of their applications to quantum annealing, universal quantum computation and studies of quantum many-body systems.
It is of practical importance for these studies to realize fast and accurate tunable coupling between KPOs in a simple manner.
We develop a simple scheme of fast tunable coupling of KPOs with high tunability in speed and amplitude using the fast transitionless rotation of a KPO in the phase space based on the shortcuts to adiabaticity.
Our scheme enables rapid switching of the effective coupling between KPOs,
and can be implemented with always-on linear coupling between KPOs, by controlling the phase \textcolor{black}{of the pump field} and \textcolor{black}{the resonance frequency of the KPO} without controlling the amplitude of the pump field nor using additional drive fields \textcolor{black}{and couplers}.
We apply the coupling scheme to a two-qubit gate, and show that our scheme realizes high gate fidelity compared to a purely adiabatic one, by mitigating undesired nonadiabatic transitions.
\end{abstract}

\maketitle

\section{Introduction}
In the mid-twentieth century, classical parametric phase-locked oscillators \cite{Onyshkevych1959,Goto1959}, called parametrons were utilized as classical bits of digital computers.
Recently, \textcolor{black}{Kerr parametric oscillators} (KPOs) \textcolor{black}{sometimes called Kerr-cat qubits}~\cite{Milburn1991,Wielinga1993,Goto2016}, which are parametrons in the single-photon Kerr regime~\cite{Wang2019,Grimm2020} where the nonlinearity is larger than the decay rate, attracted increasing attention in terms of their applications to quantum information processing~\cite{Goto2019} and studies of quantum many-body systems~\cite{Dykman2018,Rota2019}.


In the circuit-QED architecture, which is a promising platform of quantum information processing \cite{You2005,Gambetta2017,Wendin2017,Krantz2019,Gu2019,Blais2020},
KPOs can be implemented~\cite{Meaney2014,Wang2019,Goto2019,Grimm2020} by a superconducting resonator with Kerr-nonlinearity realized by the Josephson junctions, driven by an oscillating pump field.
Two coherent states with opposite phases can exist stably in a KPO, and are used as qubit states.
Bit-flip error of a KPO is suppressed because of the stability of the coherent states against photon loss,
and thus phase-flip error dominates bit-flip error in a KPO.
It is expected that quantum error correction for KPOs can be performed with less overhead owing to such biased errors compared to conventional qubits with unbiased errors~\cite{Tuckett2019,Ataides2021}.

Quantum annealing \cite{Goto2016,Nigg2017,Puri2017,Zhao2018,Onodera2020,Goto2020a,Kanao2021}
and universal quantum computation \cite{Cochrane1999,Goto2016b,Puri2017b} using KPOs have been studied theoretically.
Single-qubit operations were experimentally demonstrated~\cite{Grimm2020}.
Two-qubit gate operations, which preserve the biased feature of errors and allow one to use its advantage, were studied theoretically~\cite{Puri2020}, and 
high error-correction performance by concatenating the XZZX surface code~\cite{Ataides2021} with KPOs were numerically demonstrated~\cite{Darmawan2021}.
Fast and accurate controls~\cite{Kanao2021b,Xu2021,Kang2021}, spectroscopy~\cite{Yamaji2021,Masuda2021b}, controls and dynamics not confined in qubit space~\cite{Zhang2017,Wang2019}, Boltzmann sampling~\cite{Goto2018}, effect of strong pump field~\cite{Masuda2020}\textcolor{black}{, effect of decay and dephasing \cite{Puri2017b}}, quantum phase transitions~\cite{Dykman2018,Rota2019} and quantum chaos~\cite{Milburn1991,Hovsepyan2016,Goto2021b} have been the subject of investigations of KPO systems.
Many of the above studies use multi-KPO systems where the inter-KPO coupling plays a major role determining the property of the system.
Simple coupling scheme of KPOs with high tunability in terms of speed and amplitude will extend the degrees of freedom of controls, and is highly desirable for significant advances in the fields.

Many of the relevant control schemes of KPOs resort to quantum adiabatic dynamics~\cite{Goto2019}.
However, in practice, there are unwanted excitations due to the violation of the quantum adiabatic theorem in the controls, when performed in a short time.
Fast and accurate manipulations of KPOs have been studied using the shortcuts to adiabaticity (STA)~\cite{Rice2003,Torrontegui2013,Masuda2015,Masuda2016,Palmero2016,Campo2019,Guery-Odelin2019,Lizuain2019}, a group of protocols which mitigate or eliminate completely such unwanted excitations realizing the desired final state.
Fast creation of a cat state (a superposition of two coherent states with opposite phases) \cite{Puri2017} and traveling cat states \cite{Goto2019b} and
geometric quantum computation with cat qubits were proposed~\cite{Kang2021} based on the STA.


In this paper, we develop a scheme of fast tunable $ZZ$ coupling of KPOs using the counter-diabatic (CD) protocol~\cite{Rice2003,Masuda2016}, which is categorized to the STA.
The coupling scheme is based on a fast transitionless rotation of a KPO in the phase space, 
and importantly can be implemented with the fixed amplitude of the pump field and with always-on coupling between resonators constituting the KPOs in contrast to other schemes~\cite{Goto2016b,Puri2017b,Puri2020},
and moreover does not require additional driving fields \textcolor{black}{ in contrast to the schemes in Ref.~\cite{Darmawan2021,Chono2022}}.
In our scheme, the coupled KPOs can be identical because the controlled relative phase of the pump fields can eliminate undesired energy transfers between KPOs. Thus, the scheme will mitigate hardware requirements, complexity of sample design and frequency crowding, which are critical and ubiquitous problem of current quantum computing technologies.
We apply this scheme to $ZZ$ rotation ($R_{zz}$ gate), and show that our scheme realizes high fidelity compared to a purely adiabatic scheme,  mitigating undesired nonadiabatic transitions.


The $R_{zz}$ gate using the ideal tunable coupling, $g(t)(a_1a_2^\dagger + a_1^\dagger a_2)$ called beam-splitter type, was studied in Ref.~[\citenum{Goto2016b,Puri2017b,Puri2020}], however without the crucial analysis on the implementation of the coupling. 
($g$ and $a_l$ are the coupling amplitude and annihilation operator of resonator $l$, respectively.) 
\textcolor{black}{The tunable coupling approximating the beam-splitter type was implemented
between superconducting resonators using a transmon as a coupler~\cite{Gao2018}, and
between a KPO and a readout cavity using an additional microwave drive~\cite{Grimm2020}.}
\textcolor{black}{$R_{zz}$ gates using an additional drive to either of KPO were also proposed~\cite{Darmawan2021,Chono2022}, which require qubits with different frequencies. }
\textcolor{black}{On the other hand,
our scheme does not require couplers (KPOs can be directly coupled) nor difference between qubit  frequencies. 
}

\section{Transitionless rotation of a KPO}
Before introducing the coupling scheme, we first develop the method of the fast transitionless rotation of a KPO used for the coupling scheme.
We consider a KPO of which Hamiltonian is written in a rotating frame as~\cite{Wielinga1993,Cochrane1999,Goto2019}
\begin{eqnarray}
\frac{H(\theta)}{\hbar} = \frac{K}{2}a^{\dagger 2} a^2 - \frac{p}{2} (a^{\dagger 2}e^{2i\theta} + a^2e^{-2i\theta}), 
\label{H_1_24_22}
\end{eqnarray}
where $K$ is the nonlinearity parameter, $p$ and $2\theta$ are the amplitude and phase of the pump field (see also Appendix~\ref{Hamiltonian of a parametron2}).
Hereafter, we assume that $K$ and $p$ are positive for simplicity, although they are negative for realized KPO reported e.g. in Ref.~[\citenum{Wang2019}], because the overall sign of the Hamiltonian is not of physical importance~\cite{Goto2019}.
\textcolor{black}{The nonlinearity of the system was implemented by Josephson junctions, e.g., in Ref.~[\citenum{Meaney2014,Wang2019}]. (The intrinsic non-linearity of disordered superconductors such as granular aluminum~\cite{Winkel2020} is also expected to be used as a source of the nonlinearity.)}
This system has two degenerate ground states represented as $(|\alpha e^{i\theta} \rangle \pm |-\alpha e^{i\theta} \rangle)/\sqrt{2}$ with $\alpha=\sqrt{p/K}$ when $p\gg K$, which are called the even and odd cat states, respectively. 

\begin{figure}[h!]
\begin{center}
\includegraphics[width=8.5cm]{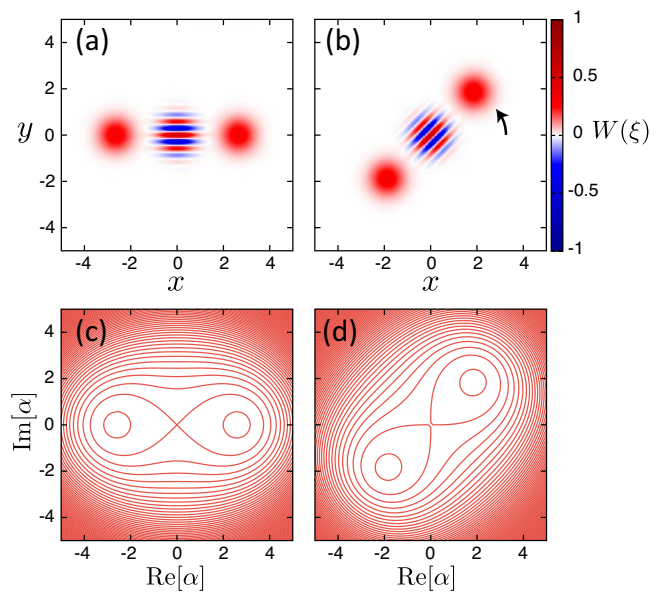}
\end{center}
\caption{
The Wigner function \textcolor{black}{$W(\xi)$} of the even cat state for $\theta=0$ (a) and $\pi/4$~(b)\textcolor{black}{, where $\xi=x+iy$}.
We set $p/K=7$.
Illustration of effective potential $V(\alpha)$ for $\theta=0$~(c) and $\pi/4$~(d).
}
\label{Wig0_11_25_21}
\end{figure}

The phase of the pump field determines the orientation of the Wigner function of energy eigenstates of the KPO~\cite{Puri2020} (see Appendix~\ref{Rotation and disturbances due to nonadiabatic transitions} for definition of the Wigner function).
Figure~\ref{Wig0_11_25_21} shows the Wigner function of the even cat state for $\theta=0$ and $\pi/4$.
As explained later, this phase dependence of KPOs can be used to tune the effective coupling between KPOs.
In order to intuitively understand the phase dependence of KPOs, we consider the effective potential defined by $V(\alpha)=\langle \alpha | H | \alpha \rangle$~\cite{Zhang2017} with a complex variable $\alpha$.
$V(\alpha)$ is represented as 
$V(\alpha)=|\alpha|^2 \Big{(} \frac{K}{2}|\alpha|^2 - p\cos(2(\theta_\alpha - \theta) ) \Big{)}$, 
where $\theta_\alpha={\rm arg}[\alpha]$.
The effective potential is oriented with the increase of $\theta$ as illustrated in 
Fig.~\ref{Wig0_11_25_21}(c) and \ref{Wig0_11_25_21}(d), and
the orientation of the effective potential coincides with the one of the Wigner function.

The Wigner function can be rotated by changing $\theta$ gradually.
When the rate of change of $\theta$ is sufficiently small, an adiabatic dynamics leads to a simple rotation of the Wigner function.
On the other hand, when the rate of change of $\theta$ is large, the Wigner function is disturbed due to unwanted nonadiabatic transitions as shown in Appendix~\ref{Rotation and disturbances due to nonadiabatic transitions}.

\textcolor{black}{Unwanted nonadiabatic transitions can be eliminated by adding the detuning of which time dependence is designed by the CD protocol~\cite{Rice2003}.
The modified Hamiltonian is represented as
\begin{eqnarray}
H'(t) = H(\theta(t)) - \hbar \dot\theta(t) a^\dagger a,  
\end{eqnarray}
where dot denotes the time derivative.
The modified Hamiltonian $H'$ is composed of $H(\theta)$ in Eq.~(\ref{H_1_24_22}) and $- \hbar \dot\theta(t) a^\dagger a$, which we call CD term (see Appendix~\ref{Theory of rotation} for the derivation of the CD term).
The additional detuning to eliminate nonadiabatic transitions during the rotation was obtained for constant rate of the rotation for a similar system~\cite{Guillaud2019}. And the same detuning term appears when the Hamiltonian for the transitionless CX gate when projected to a particular state of the control qubit~\cite{Puri2020}. }


The CD term can be implemented by the detuning $\Delta$ in KPOs, which is the difference between the resonance frequency of the KPO and half of the pump frequency~\cite{Goto2016,Goto2019} and appears as a term, $\hbar\Delta a^\dagger a$, in the KPO Hamiltonian~\cite{Goto2019}.
The detuning can be tuned by controlling the resonance frequency of the KPO via the magnetic flux~\cite{Wang2019}.
\textcolor{black}{(It is known that the resonance frequency of a superconducting resonator can be modulated half a gigahertz in 1~ns~\cite{ZLWang2013}.)}
\textcolor{black}{The resonance frequency of the KPO is modified as $\omega(t) = - \Delta(t)  + K + \omega_p/2$, where $\omega_p$ is the angular frequency of the pump field (see Appendix~\ref{Hamiltonian of a parametron2} for details).} 
Therefore, controlling the phase of the pump field \textcolor{black}{and the resonance frequency of the KPO} can rotate a KPO without any disturbance.
The performances of the controls with and without the CD term are compared by numerical simulations in Appendix~\ref{Rotation and disturbances due to nonadiabatic transitions}.

\section{Fast tunable coupling scheme}
We introduce a scheme of fast tunable coupling for KPOs based on the above transitionless rotation of a single KPO. 
We consider two linearly coupled KPOs with the same resonance frequencies with Hamiltonian 
\begin{eqnarray}
\frac{H_{\rm tot}(t)}{\hbar} &=&  \sum_{l=1}^{2} \Big{[} \frac{K_l}{2}a_l^{\dagger 2} a_l^2 - \frac{p_l}{2} (a_l^{\dagger 2}e^{{2i\theta_l(t)}} + a_l^2e^{{-2i{\theta_l(t)}}}) \Big{]} \nonumber\\
&& + J (a_1a_2^\dagger + a_1^\dagger a_2)
-  \dot\theta_1(t) a_1^\dagger a_1,
\label{H_2KPO_11_25_21}
\end{eqnarray}
where $K_l$ is the nonlinearity parameter, $p_l$ and $2\theta_l$ are the amplitude and phase of the pump field of KPO $l$. Here, $J$ is the fixed coupling coefficient between the KPOs.
We emphasize that the effective coupling between KPOs can be turned off even with fixed $J$ as shown below.
The last term in Eq.~(\ref{H_2KPO_11_25_21}) is the CD term for transitionless rotation of KPO 1.
We, hereafter, assume that $p_l=p$, $K_l=K$ and $\theta_1(t)=\theta(t)$ and $\theta_2=0$ for simplicity.
Note that only the phase of KPO 1 is modulated, while that of KPO 2 is fixed.
\textcolor{black}{We assume that the phase of the pump fields can be precisely controlled in this study.
There might be slow changes of the phase due to phase drifts in actual experiments. Such phase drift can be monitored during the measurement, and the phase can be adjusted before each measurement~\cite{Wang2019}.}

In the parameter regime where $p\gg J$ and $K$,
four states represented by $|\alpha e^{i\theta},\alpha \rangle$, $|\alpha e^{i\theta},-\alpha \rangle$, $|-\alpha e^{i\theta},\alpha \rangle$, $|-\alpha e^{i\theta},-\alpha \rangle$ with $\alpha=\sqrt{p/K}$ are stable due to the exponential suppression of bit-flip rate caused with the increase of $\alpha$ \cite{Puri2019}.
Hereafter, these states are denoted by $|\bar{0},\bar{0}\rangle$, $|\bar{0},\bar{1}\rangle$, $|\bar{1},\bar{0}\rangle$ and $|\bar{1},\bar{1}\rangle$, respectively.
The interaction terms in the Hamiltonian shift the energy of the states because 
\begin{eqnarray}
\langle \bar{0},\bar{0} (\bar{1},\bar{1})| (a_1 a_2^\dagger + a_1^\dagger a_2) |  \bar{0},\bar{0} (\bar{1},\bar{1}) \rangle
&=& 2|\alpha|^2 \cos\theta, \nonumber\\
\langle \bar{0},\bar{1} (\bar{1},\bar{0}) | (a_1 a_2^\dagger + a_1^\dagger a_2) |  \bar{0},\bar{1} (\bar{1},\bar{0}) \rangle
&=& - 2|\alpha|^2 \cos\theta 
\label{Int_11_29_21}
\end{eqnarray}
while off-diagonal elements, such as 
$\langle \bar{0},  \bar{0}  | a_1 a_2^\dagger + a_1^\dagger a_2 |  \bar{0},  \bar{1} \rangle$, are negligible (note that $\langle -\alpha| \alpha \rangle\simeq 0$).
Importantly, the shift of the energies can be controlled via $\theta$ as seen in Eq.~(\ref{Int_11_29_21}), and the shift of the energies becomes zero when $\theta = \pi/2 + \pi n$, where $n$ is an integer.
Thus, the effective coupling between KPOs can be tuned and even turned off. 
\textcolor{black}{(This controllability of the effective coupling is not lost even in systems with asymmetry between KPOs, as explained in Appendix~\ref{Asymmetry}.)}
A pulsed effective coupling can be used to perform a $R_{zz}$ gate as shown bellow. 

We consider the phase of the pump field
controlled for $0 \le t \le T$ as $\theta(t) = \frac{\pi}{2} - \theta_{\rm amp} \pi [1-\cos(2\pi t/T)]$,
where $\theta_{\rm amp}$ is a constant parameter, which determines maximum strength of the effective coupling during the control.
$\theta$ is chosen to be $\pi/2$ at $t=0$ and $T$ so that the effective coupling is off at the initial and final times of the control. 
We assume that the initial state is one of the states, $|\bar{0},\bar{0}\rangle$, $|\bar{0},\bar{1}\rangle$, $|\bar{1},\bar{0}\rangle$ and $|\bar{1},\bar{1}\rangle$.
For sufficiently large $T$, the state of the system evolves adiabatically from $|i,j\rangle$ to $e^{i\varphi_{ij}}|i,j\rangle$, where $i,j=\bar{0},\bar{1}$.
Here, $\varphi_{ij}$ is the dynamical phase at $t=T$ due to the energy shift, and is written 
\begin{eqnarray}
\varphi_{ij} = \left\{
\begin{array}{cl}
2J|\alpha|^2\int_0^T  \cos\theta(t) dt &  {\rm for} \ i=j, \\
-2J|\alpha|^2\int_0^T \cos\theta(t) dt & {\rm for} \ i\ne j
\end{array}
\right.
\label{theta_1_6_22}
\end{eqnarray}
\textcolor{black}{(see Appendix~\ref{Asymmetry} for the case that there is asymmetry between the KPOs)}.
Thus, we can perform $R_{zz}$ gates simply by controlling the phase of the pump field.
\textcolor{black}{The dynamical phase increases with the rate of $2J|\alpha|^2\cos\theta(t)$.
The maximum value of $|\varphi_{ij}|$ is $2J|\alpha|^2T$.
In other word, the minimum duration for $\varphi_{ij}$ to be realized is given by $T=|\varphi_{ij}|/2J|\alpha|^2$.}
When $T$ is sufficiently large, the CD term is not necessary because it is
proportional to $\dot{\theta}$, and therefore 
 much smaller than the other parameters.
However, for the small $T$ regime where $\dot\theta$ is comparable to or greater than the other parameters, the final state considerably deviates from $e^{i\varphi_{ij}}|i,j\rangle$ due to nonadiabatic transitions unless the CD term is used.

\begin{figure}[h!]
\begin{center}
\includegraphics[width=7cm]{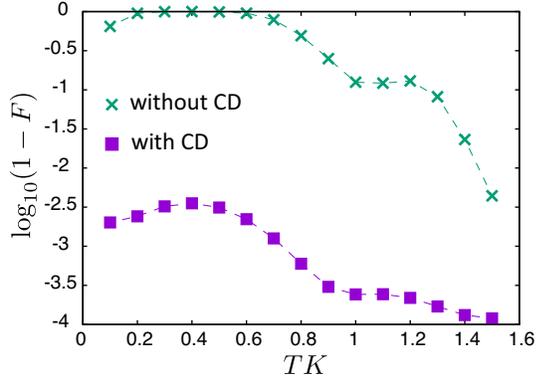}
\end{center}
\caption{
\textcolor{black}{Dependence of the infidelity on $T$ for the controls with and without the CD term for the initial state of $|\Psi_s\rangle$.
The used parameters are $p/K=7$, $J/K=0.2$ and $\theta_{\rm amp}=0.1$.
The dashed lines are guides to eyes.
}
}
\label{fid_T_1_12_22}
\end{figure}

In order to compare the performance of the controls with and without the CD term,
we numerically simulate the dynamics with the initial state of $|\Psi(0)\rangle=|i,j\rangle$ and $|\Psi_s\rangle\equiv\sum_{i,j}|i,j\rangle/\sqrt{4}$, and obtain the fidelity  defined by \textcolor{black}{$F=$}$|\langle \Psi_{\rm ideal} | \Psi(T)\rangle|^2$, where $|\Psi_{\rm ideal}\rangle=U_{\rm ideal}|\Psi(0)\rangle$, and $U_{\rm ideal}$ denotes the operator representing the ideal gate operation, e.g., $U_{\rm ideal}|i,j\rangle =  e^{i\varphi_{ij}}|i,j\rangle$.
Figure~\ref{fid_T_1_12_22} shows the dependence of the \textcolor{black}{infidelity} on $T$ for the controls with and without the CD term for $\theta_{\rm amp}=0.1$ \textcolor{black}{for the initial state of $|\Psi_s\rangle$}. 
\textcolor{black}{ The results for the initial state of $|i,j\rangle$ are approximately the same as the ones for the initial state of $|\Psi_s\rangle$.}
The fidelity of the control without the CD term (purely adiabatic scheme) is degraded by the nonadiabatic transitions as $T$ decreases. 
On the other hand, the fidelity of the control with the CD term is approximately unity in such small $T$ regime.
For example, the fidelity of the control with the CD term averaged over the initial states is approximately 0.9995, while the averaged fidelity of the control without the CD term is less than 0.89 for $T=K^{-1}$.
The fidelity of the control with the CD term slightly decreases from unity as $T$ decreases. 
We attribute this to the fact that the CD term is designed for transitionless rotation of an individual KPO ($J=0$), and therefore there is finite nonadiabatic transitions for $J\ne 0$.
However, it is noteworthy that the CD term can work well also for the case with $J\ne 0$.
\textcolor{black}{
The infidelity for both the controls decreases in the short-$T$ regime. 
We attribute this to the followings: the control duration is so short that the deviation of the final state of the KPO from its initial state is moderate; the phase in Eq.~(\ref{theta_1_6_22}) to be imprinted on the state of the KPO at $t=T$ is small for short $T$, and therefore the target state $|\Psi_{\rm ideal}\rangle$ is approximately the same as the initial state.}

\begin{figure}[h!]
\begin{center}
\includegraphics[width=7cm]{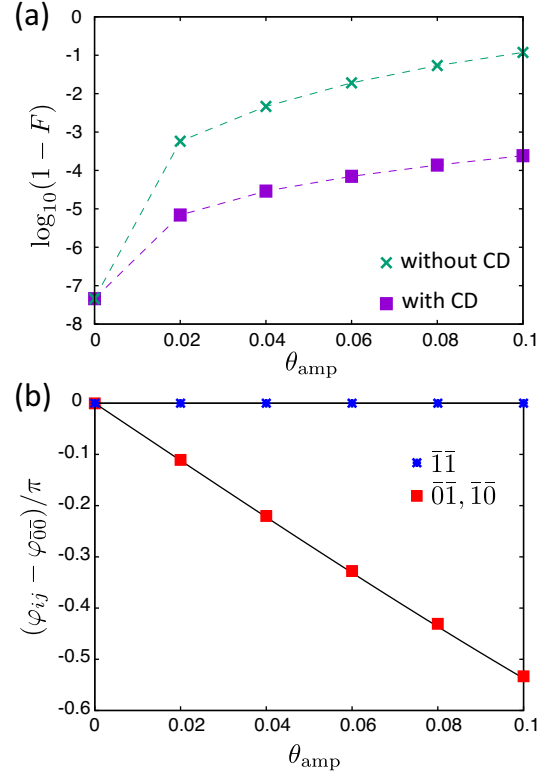}
\end{center}
\caption{
(a) Dependence of the \textcolor{black}{infidelity} on $\theta_{\rm amp}$ for the controls with and without the CD term \textcolor{black}{for the initial state of $|\Psi_s\rangle$}. 
The dashed lines are guides to eyes.
(b) Dependence of $\varphi_{ij}$ on $\theta_{\rm amp}$ of the control with the CD term.
The solid curve represents $\varphi_{\bar{0}\bar{1}(\bar{1}\bar{0})}$ in Eq.~{(\ref{theta_1_6_22})}.
The used parameters are $p/K=7$, $J/K=0.2$ and $T=K^{-1}$.
}
\label{fid_sft1_11_30_21}
\end{figure}

Figure~\ref{fid_sft1_11_30_21}(a) shows the dependence of the \textcolor{black}{infidelity} on $\theta_{\rm amp}$ for the controls with $T=K^{-1}$ \textcolor{black}{for the initial state of $|\Psi_s\rangle$}. 
\textcolor{black}{ The results for the initial state of $|i,j\rangle$ are approximately the same as the ones for the initial state of $|\Psi_s\rangle$.}
It is seen that the fidelity of the control with the CD term is much higher than that of the control without the CD term for large $\theta_{\rm amp}$.
\textcolor{black}{The fidelity of the control with the CD term decreases, as well as the control without the CD term, with the increase of $\theta_{\rm amp}$. This is attributed to the fact that the CD term is exact only for the case of $J=0$, and that nonadiabatic transitions increase with $\theta_{\rm amp}$.}
We numerically obtain phase $\varphi_{ij}$, which the system acquires during the control with the CD term, and compare it with the analytic one in Eq.~(\ref{theta_1_6_22}).
We define the phase at $t=T$ as $\varphi_{ij} = \arg [ \langle ij | \Psi(T)\rangle ]$
for the simulation with the initial state, $| \Psi(0)\rangle = |i,j\rangle$. 
Figure~\ref{fid_sft1_11_30_21}(b) shows the dependence of $\varphi_{ij}$ on $\theta_{\rm amp}$ for $T=K^{-1}$.
The relative phases, $\varphi_{\bar{1}\bar{0}}-\varphi_{\bar{0}\bar{0}}$ and $\varphi_{\bar{0}\bar{1}}-\varphi_{\bar{0}\bar{0}}$, monotonically decrease with $\theta_{\rm amp}$ in the used range of $\theta_{\rm amp}$, while $\varphi_{\bar{1}\bar{1}}-\varphi_{\bar{0}\bar{0}}$ is approximately zero.
It is seen that numerical results agree well with Eq.~(\ref{theta_1_6_22}).



\begin{figure}[h!]
\begin{center}
\includegraphics[width=7cm]{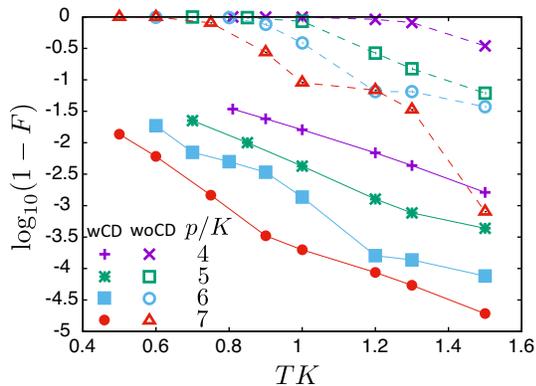}
\end{center}
\caption{
\textcolor{black}{
Dependence of the infidelity on $T$ for the controls with and without the CD term 
for various values of $p$.
The initial state is $|\Psi_s\rangle$ and $J/K=0.2$.
The value of  $\theta_{\rm amp}$ was chosen so that  $(\varphi_{\bar{1}\bar{0}} - \varphi_{\bar{0}\bar{0}})/\pi=-0.5$.
The solid and dashed lines are guides to eyes.
}
}
\label{fid_fin_com_6_18_22_2}
\end{figure}

\textcolor{black}{Figure~\ref{fid_fin_com_6_18_22_2} shows the $T$-dependence of the infidelity 
of $R_{zz}$ gate with $(\varphi_{\bar{1}\bar{0}} - \varphi_{\bar{0}\bar{0}})/\pi=-0.5$ for various amplitude of the pump field, $p$.
The value of  $\theta_{\rm amp}$ was chosen so that $(\varphi_{\bar{1}\bar{0}} - \varphi_{\bar{0}\bar{0}})/\pi=-0.5$.
It is seen that the CD term considerably increases the fidelity in the parameter range studied.
The fidelity also increases with respect to $p$ because the nonadiabatic transitions are mitigated for larger pump amplitude due to the increase of the gap between energy levels~\cite{Masuda2020}.}
\textcolor{black}{The performance of our coupling scheme is compared with the one based on the ideal tunable coupling with the form of $g(t)(a_1a_2^\dagger + a_1^\dagger a_2)$ in Appendix~\ref{Beam splitter type}.}

\textcolor{black}{
As exemplified in Fig.~\ref{fid_fin_com_6_18_22_2}, the gate fidelity can be increased by increasing control duration $T$.
However, the gate fidelity is decreased  for larger $T$ when there is the decoherence.
We examine the performance of the controls under the effect of the decoherence using the master equation:
\begin{eqnarray}
\frac{d\rho(t)}{dt} &=& -\frac{i}{\hbar}[H_{\rm tot}(t),\rho(t)] + \mathcal{L}[\rho(t)],\nonumber\\
\mathcal{L}[\rho] &=& \sum_l \frac{\kappa_l}{2} ([a_l\rho,a_l^\dagger] + [a_l,\rho a_l^\dagger])\nonumber\\
&&+\gamma_p^{(l)} ([a_l^\dagger a_l\rho,a_l^\dagger a_l] + [a_l^\dagger a_l,\rho a_l^\dagger a_l]), 
\label{ME_6_28_22}
\end{eqnarray}
where $\kappa_l$ and $\gamma_p^{(l)}$ are the decay and dephasing rates of KPO $l$.
We assume that $\gamma_p^{(l)},\kappa_l=\kappa$ in numerical simulations for simplicity (the effect of  the pure decay and pure dephasing are examined in Appendix~\ref{Robustness}).
Figure~\ref{fid_fin_com_kappa_6_18_22} shows the dependence of the infidelity on $T$ for $\kappa=10^{-3}K$ and $10^{-4}K$.
It is seen that the fidelity of the control with the CD term is greatly improved in the short-$T$ regime compared to that of the control without CD term, while the efficiency of both the controls are degraded for large $T$.
This result shows that the control with the CD term allows fast $R_{zz}$ gates with high fidelity avoiding unwanted effects of the decoherence and nonadiabatic transitions.
}
\textcolor{black}{
At the minimums and a plateau of the infidelity seen in Fig.~\ref{fid_fin_com_kappa_6_18_22}, the mitigation of nonadibatic transitions 
accompanied with the increase of $T$ is balancing with the effect of decoherence.
}

\begin{figure}[h!]
\begin{center}
\includegraphics[width=7cm]{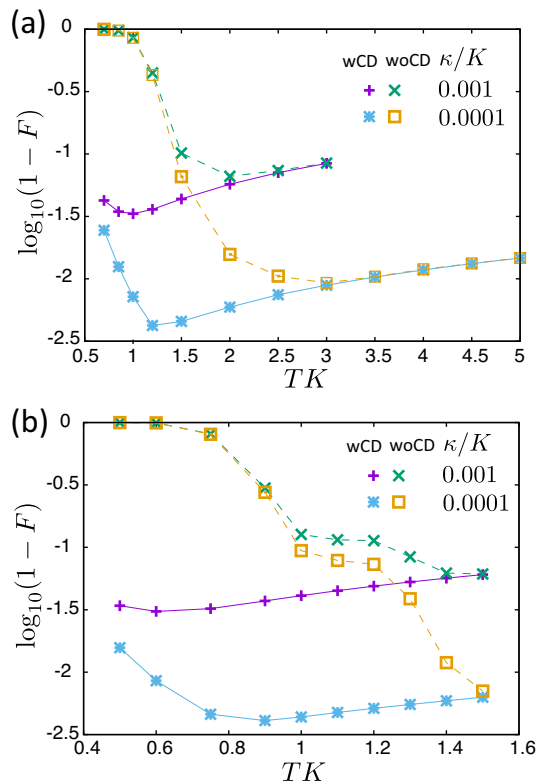}
\end{center}
\caption{
\textcolor{black}{
Dependence of the infidelity on $T$ for the controls with and without the CD term 
for $J/K=0.2$, $\kappa=10^{-3}K$ and $10^{-4}K$ for $p/K=5$ (a) and  $p/K=7$ (b).
The initial state is $|\Psi_s\rangle$.
The value of  $\theta_{\rm amp}$ was chosen so that  $(\varphi_{\bar{1}\bar{0}} - \varphi_{\bar{0}\bar{0}})/\pi=-0.5$.
The solid and dashed lines are guides to eyes.
}
}
\label{fid_fin_com_kappa_6_18_22}
\end{figure}

\textcolor{black}{
Because the detuning of a KPO opens the energy gap between energy levels~\cite{Goto2016b}, imperfection of the control of the detuning can disturb the state of the KPO.  
The robustness of our scheme against the error in the CD term is examined in Appendix~\ref{Robustness}.}

\section{Conclusions and discussions}
A fast tunable $ZZ$ coupling scheme of KPOs has been developed using the transitionless rotation of a KPO in the phase space based on the CD protocol.
The effective coupling between KPOs can be turned off even with always-on linear coupling between the resonators constituting the KPOs. 
We have examined the performance of our scheme applying it to $R_{zz}$ gate, and compared with the results of a purely adiabatic scheme, which utilizes only a controlled phase of the pump field.
It has been shown that our scheme greatly enhances the fidelity of $R_{zz}$ gate compared to the adiabatic scheme by eliminating undesired nonadiabatic transitions, when applied in a short time.

\textcolor{black}{
The CD protocol can be realized by the time dependent detuning implemented by controlling the resonance frequency of the KPO. 
Because time dependent detuning can be used also for the $R_x$ gate~\cite{Goto2016b}, our scheme is compatible with the $R_x$ gate in the sense that the use of the time dependent detuning will not add extra experimental equipment and tasks, such as calibration, to complete the set of universal gates. }

\textcolor{black}{A KPO can be loaded to its ground state, which is a cat state, from the vacuum state by adiabatically ramping the pump amplitude~\cite{Cochrane1999,Goto2016}.
We examine the efficiency of the adiabatic loading scheme when the effective coupling between the KPOs is off in Appendix~\ref{Loading into ground state}.}

While we are preparing our manuscript, we came to know that other group independently studied the tunability of the effective coupling solely by the phase of the pump field~\cite{NEC_paper}. 
However, this method has recourse to an adiabatic evolution of the system and, therefore, is not suitable for fast tuning of the coupling. Our scheme resolves the shortcoming of the adiabatic scheme.

\textcolor{black}{A comment on the readout is in order. 
We assume that each KPO is read using the output field from the KPO through a different readout transmission line. When the coupling between the KPOs are off, the state of one KPO is a superposition of $|\alpha\rangle$ and $|-\alpha\rangle$ while that of the other is a superposition of $|i\alpha\rangle$ and $|-i\alpha\rangle$. Even if there is a small leakage from KPO~1 to the readout transmission line for KPO~2, they can be distinguished because of the phase difference of $\pi/2$.}
\textcolor{black}{Reading out a KPO by coupling it to a transmission line may shorten the lifetime of the cat states. Such unwanted effect will be mitigated by using a readout cavity attached to the KPO~\cite{Grimm2020}.
}

Before closing, we point out that our coupling scheme will find wider applications in quantum technologies, although we particularly applied to $R_{zz}$ gate in this paper to demonstrate the effectiveness of the scheme.
For example, the coupling scheme can be useful for quantum annealing and quantum simulation in which time dependent qubit-qubit couplings are utilized.
Our scheme is used to decrease undesired population transfers out of the qubit space caused by the rotation of a KPO (not nonadiabatic transitions of the whole system which may be caused by time dependent effective coupling), and therefore has a different motivation from other studies based on the STA which consider ideal spin models and aim decreasing nonadiabatic transitions of the model systems~\cite{delCampo2012,Damski2014,Okuyama2016,Sels2017,Setiawan2019}.
Our scheme can be implemented by the simple manner and even independent of energy-level structure of the system. Performance of our scheme in quantum annealing and quantum simulation deserves further quantitative investigation.

\begin{acknowledgments}
It is a pleasure to acknowledge discussions with T. Yamamoto.
This paper is partly based on results obtained from a project, JPNP16007, commissioned by the New Energy and Industrial Technology Development Organization (NEDO), Japan. 
S.M. acknowledges the support from JSPS KAKENHI (grant number 18K03486). 
Y. M. was supported by MEXT's Leading Initiative for Excellent Young Researchers and JST PRESTO (Grant No. JPMJPR1919), Japan.
\end{acknowledgments}

\appendix

\begin{figure}
\begin{center}
\includegraphics[width=4.5cm]{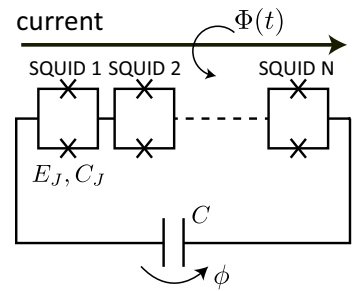}
\end{center}
\caption{
\textcolor{black}{
Circuit model of a KPO consisting of $N$~SQUIDs and a shunt capacitor $C$. 
$\phi$ is the overall phase across the junction array;  $\Phi(t)$ is the external magnetic flux threading the SQUIDs; $E_J$ and $C_J$ are the Josephson energy of a single SQUID and the capacitance of a single Josephson junction, respectively.  
}
}
\label{KPO_system_6_1_20}
\end{figure}

\textcolor{black}{
\section{Hamiltonian of a KPO}
\label{Hamiltonian of a parametron2}}
\textcolor{black}{
Although a derivation of an effective Hamiltonian for a KPO was shown in Ref.~\cite{Wang2019},
we present it to make this paper self-contained.
We consider a KPO composed of a SQUID-array resonator with $N$ SQUIDs illustrated in  Fig.~\ref{KPO_system_6_1_20}.
The effective Hamiltonian of the system is represented as
\begin{eqnarray}
H= 4E_C n^2 - NE_J[\Phi(t)] \cos\frac{\phi}{N},
\label{H_KPO_4_16_20}
\end{eqnarray}
where $\phi$, $n$, $E_J$ and $E_C$ are the overall phase across the junction array, its conjugate variable and the Josephson energy of a SQUID, respectively. 
$E_C$ is the charging energy of the resonator, including the contributions of the junction capacitances $C_J$ and the shunt capacitance $C$.
We assume that all the Josephson junctions are identical.
The effective Hamiltonian (\ref{H_KPO_4_16_20}) with a single degree of freedom, $\phi$, is valid provided that $E_J$ is much larger than the charging energy of a single junction~\cite{Frattini2017,Noguchi2020}.
The Josephson energy can be modulated as $E_J(t)=E_J+\delta E_J \cos\omega_p t$ by the time-dependent external magnetic flux, $\Phi(t)$, threading the SQUIDs.
For simplicity, we set the phase of the pump field to be zero, $\theta=0$.
}

\textcolor{black}{
We can obtain an approximate Hamiltonian by taking into account up to the fourth order of $\phi/N$ in equation~(\ref{H_KPO_4_16_20}) as
\begin{eqnarray}
\frac{H}{\hbar} &=& \omega \Big{(} a^\dagger a + \frac{1}{2} \Big{)}
- \frac{K}{12} (a + a^\dagger)^4
\nonumber\\
 && + \Big{[} - \frac{N\delta E_J}{\hbar}  +
 p (a + a^\dagger)^2  - \frac{K p}{3\omega} (a + a^\dagger)^4 \Big{]}\nonumber\\
 && \times \cos\omega_p t,
\label{H_9_1_20}
\end{eqnarray}
where $\omega = \frac{1}{\hbar}\sqrt{8E_CE_J/N}$, $K=E_C/\hbar N^2$ and $p = 2\omega \delta E_J/ 8E_J$. 
$n$ and $\phi$ are related to the annihilation operator $a$ as
$n = -in_0(a-a^\dagger)$ and $\phi = \phi_0 (a+a^\dagger)$
with $n_0^2=\sqrt{E_J/32 N E_C}$ and $\phi_0^2 = \sqrt{2NE_C/E_J}$.
Above, we considered the parameter regime, where $\phi_0/N = 2\sqrt{K/\omega}$ is sufficiently smaller than unity so that the expansion of $\cos(\phi/N)$ is valid,
and took into account up to the fourth order of $\phi/N$ to see the effect of the Kerr nonlinearity.
In equation~(\ref{H_9_1_20}), we neglect the last term assuming that $Kp\ll  \omega$, and drop c-valued terms to obtain the following Hamiltonian
\begin{eqnarray}
\frac{H}{\hbar} = \omega a^\dagger a 
- \frac{K}{12} (a + a^\dagger)^4
+ p (a + a^\dagger)^2
\cos\omega_p t.
\end{eqnarray}
Moving to the rotating frame at the frequency of $\omega_p/2$ and using the rotating wave approximation, we obtain 
\begin{eqnarray}
\frac{H}{\hbar} = -\Delta a^\dagger a - \frac{K}{2}a^{\dagger 2} a^2 + \frac{p}{2}(a^{\dagger 2} + a^2),
\end{eqnarray}
where $\Delta=-\omega + K + \omega_p/2$.
Puting $\Delta=0$ and changing the sign of the Hamiltonian we can obtain Eq.~(\ref{H_1_24_22}) for $\theta=0$.
}
\textcolor{black}{When we take into account higher order terms with respect to $\phi$, we have higher order terms with respect to $a$ such as the term proportional to  $\frac{pK}{\omega}a^\dagger a^3$.
The higher order terms can be neglected when $p$ and $K$ are much smaller than $\omega$. We consider such parameter regimes throughout this paper.
We also refer readers to Ref.~[\citenum{Puri2017b}], which studied the effect of higher-order terms.}

\section{Performance of rotation schemes for single KPO}
\label{Rotation and disturbances due to nonadiabatic transitions}
We compare the performance of the rotation schemes with and without the CD term.
As an example, we consider the case that $\theta$ is increased from 0 to $\pi/2$ for $0\le t\le T$ as
\begin{eqnarray}
\theta(t) = \frac{\pi}{4}\Big{[} 1- \cos\Big{(} \frac{\pi t}{T} \Big{)} \Big{]}.
\label{theta0_11_29_21}
\end{eqnarray}
The initial state is a ground state well approximated by $(|\alpha\rangle + |-\alpha\rangle)/\sqrt{2}$, where $\alpha=\sqrt{p/K}$. 
The Wigner function of the initial state is presented in Fig.~\ref{Wig0_11_25_21}(a).
The Wigner function is defined by $W(\xi)=\frac{2}{\pi}{\rm Tr}[D(-\xi)\rho D(\xi)P]$, with $\xi=x+iy$, density operator $\rho$, displacement operator $D(\xi)=\exp(\xi a^\dagger - \xi^\ast a)$ and parity operator $P={\rm exp}(i\pi a^\dagger a)$~\cite{Leonhardt1997,Deleglise2008,Goto2016}.
We fix $p$ and $K$, while $\theta$ is changed during the control. 

\begin{figure}[h!]
\begin{center}
\includegraphics[width=8cm]{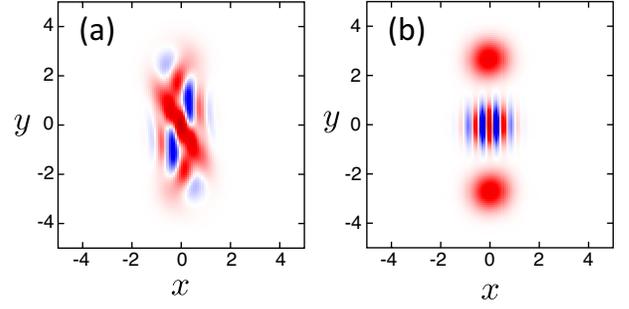}
\end{center}
\caption{
The Wigner function of the final state of the control without the CD term
for $T=0.6K^{-1}$ (b) and $T=1.5K^{-1}$.
The used parameters are the same as Fig.~\ref{Wig0_11_25_21}.
}
\label{Wig_11_25_21}
\end{figure}

Figures~\ref{Wig_11_25_21}(a) and \ref{Wig_11_25_21}(b) show the Wigner function at $t=T$ for $T=0.6K^{-1}$ and $1.5K^{-1}$, respectively, for the control without the CD term.
The Wigner function at $t=T$ is disturbed due to nonadiabatic transitions for $T=0.6K^{-1}$, while
the Wigner function is almost ideally rotated for $1.5K^{-1}$ because the system evolves almost adiabatically.

\begin{figure}[h!]
\begin{center}
\includegraphics[width=7cm]{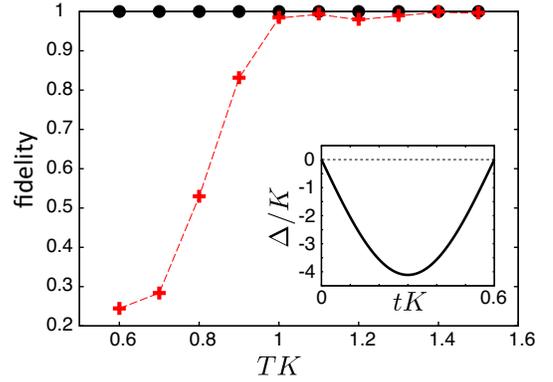}
\end{center}
\caption{
$T$-dependence of the fidelity of a rotation of a KPO.
The black circles and red crosses correspond to the controls with and without the CD term.
The inset shows the time dependence of the detuning $\Delta(t) = -\dot\theta(t)$ for $T=0.6K^{-1}$ in the control with the CD term.
}
\label{fid_com_11_28_21}
\end{figure}

We define the fidelity of the control as $|\langle\Psi_{\theta(T)}|\Psi(T)\rangle|^2$, where $|\Psi(T)\rangle$ and $|\Psi_{\theta(T)}\rangle$ are the final state of the control and the state ideally rotated by angle $\theta(T)$, respectively.
Figure~\ref{fid_com_11_28_21} shows the $T$-dependence of the fidelity for both the controls.
In the control without the CD term, the fidelity is degraded due to nonadiabatic transitions for small $T$, while the fidelity becomes close to unity for sufficiently large $T$.
On the other hand, the fidelity of the control with the CD term is unity.
The inset of Fig.~\ref{fid_com_11_28_21} shows the time dependence of the detuning $\Delta(t) = -\dot\theta(t)$ for $T=0.6K^{-1}$.

\textcolor{black}{
\section{Derivation of CD term for rotation of a KPO in phase space}
\label{Theory of rotation}
}
The rotation of a KPO is characterized by operator $U$ defined by
\begin{eqnarray}
U(\theta) = e^{i\theta a^\dagger a}.
\label{U_11_29_21_2}
\end{eqnarray}
$U(\theta)$ rotates a state of a KPO in the $\alpha$ space.
This fact is easily confirmed by letting $U$ act on coherent state $|\alpha\rangle$ to obtain
\begin{eqnarray}
a U(\theta) |\alpha\rangle = \alpha e^{i\theta} U(\theta) |\alpha\rangle,
\end{eqnarray}
where we used $U^\dagger(\theta) a U(\theta)=a e^{i\theta}$.

Suppose that $|\phi_m\rangle$ is $m$th eigenstate of $H(0)$ with eigenenergy $E_m$.
The time independent Schr\"{o}dinger equation is written as
\begin{eqnarray}
H(0) |\phi_m\rangle = E_m|\phi_m\rangle.
\end{eqnarray}
Then, we can obtain
\begin{eqnarray}
H(\theta) U(\theta) |\phi_m\rangle = E_m U(\theta) |\phi_m\rangle,
\end{eqnarray}
where we used 
\begin{eqnarray}
H(\theta) = U(\theta)  H({0}) U^\dagger(\theta).
\label{Htheta_11_29_21}
\end{eqnarray}
The above discussion shows that if $|\phi_m\rangle$ is an eigenstate of $H(0)$,
$U(\theta)|\phi_m\rangle$, which is a rotated state by $\theta$, is an eigenstate of $H(\theta)$.
This fact is independent of energy eigenstates.
Therefore, we can rotate an arbitrary state of a KPO by adiabatically changing $\theta$.

\textcolor{black}{Now we derive a modified Hamiltonian, which realizes an ideal rotation without nonadiabatic transitions, using the CD protocol~\cite{Rice2003}. 
We consider a dynamics in the system with $\theta=0$ as a reference.
Suppose that $|\Psi(t)\rangle$ is a solution of the Schr\"{o}dinger equation
\begin{eqnarray}
i\hbar \frac{d}{dt} |{\Psi}(t)\rangle = H(0)|\Psi(t)\rangle,
\label{SE_11_29_21}
\end{eqnarray}
where $H(0)$ denotes $H(\theta=0)$.
The state rotated by $\theta(t)$ is represented as $U(\theta(t))|\Psi(t)\rangle$.
We can straightforwardly obtain the relation
\begin{eqnarray}
i\hbar\frac{d}{dt} \big\{ U(\theta(t))|\Psi(t)\rangle \big\} = H' U(\theta(t)) |\Psi(t)\rangle,
\end{eqnarray}
with
\begin{eqnarray}
H' = H(\theta(t)) - \hbar \dot\theta(t) a^\dagger a,   
\end{eqnarray}
where we have used Eqs. (\ref{U_11_29_21_2}), (\ref{SE_11_29_21}), (\ref{Htheta_11_29_21}) and $U^\dagger (\theta(t)) U (\theta(t)) =1$.
The rotated state $U(\theta(t))|\Psi(t)\rangle$ is a solution of the Schr\"{o}dinger equation corresponding to Hamiltonian $H'$ composed of $H(\theta)$ in Eq.~(\ref{H_1_24_22}) and $- \hbar \dot\theta(t) a^\dagger a$, which we call CD term.
}

\textcolor{black}{
\section{Asymmetries in system}
\label{Asymmetry}
}
\textcolor{black}{We consider the case that the two KPOs are not identical.
Although the amplitude and phase of the pump fields can be externally tuned, it is difficult to exactly set the value of $K_l$ due to imperfections of the fabrication.
In order to examine the effect of the asymmetry in $K_l$, we set $K_1=K$, $K_2=rK$, $p_1=p_2=p$, where
$r$ is the constant parameter characterizing the asymmetry in $K_l$.}

\textcolor{black}{
The four states, represented by $|\alpha_1 e^{i\theta},\alpha_2 \rangle$, $|\alpha_1 e^{i\theta},-\alpha_2 \rangle$, $|-\alpha_1 e^{i\theta},\alpha_2 \rangle$, $|-\alpha_1 e^{i\theta},-\alpha_2 \rangle$ with $\alpha_1=\sqrt{p/K}$ and $\alpha_2=\sqrt{p/rK}$, are stable due to the exponential suppression of bit-flip rate when $\alpha_l$ is sufficiently large.
These states are denoted by $|\bar{0},\bar{0}\rangle$, $|\bar{0},\bar{1}\rangle$, $|\bar{1},\bar{0}\rangle$ and $|\bar{1},\bar{1}\rangle$, respectively.
Then, the counterpart of Eq.~(\ref{Int_11_29_21}) is represented as 
\begin{eqnarray}
\langle \bar{0},\bar{0} (\bar{1},\bar{1})| (a_1 a_2^\dagger + a_1^\dagger a_2) |  \bar{0},\bar{0} (\bar{1},\bar{1}) \rangle
&=& \frac{2p}{\sqrt{r}K} \cos\theta, \nonumber\\
\langle \bar{0},\bar{1} (\bar{1},\bar{0}) | (a_1 a_2^\dagger + a_1^\dagger a_2) |  \bar{0},\bar{1} (\bar{1},\bar{0}) \rangle
&=& - \frac{2p}{\sqrt{r}K} \cos\theta,
\nonumber\\
\end{eqnarray}
and off-diagonal elements, such as 
$\langle \bar{0},  \bar{0}  | a_1 a_2^\dagger + a_1^\dagger a_2 |  \bar{0},  \bar{1} \rangle$, are negligible when $\langle -\alpha| \alpha \rangle\simeq 0$.
Therefore, the coupling can be tuned via $\theta$ as in the case of identical KPOs.
The dynamical phase imprinted on these states at $t=T$, is written as
\begin{eqnarray}
\varphi_{ij} = \left\{
\begin{array}{cl}
\frac{2Jp}{\sqrt{r}K}\int_0^T  \cos\theta(t) dt &  {\rm for} \ i=j, \\
-\frac{2Jp}{\sqrt{r}K}\int_0^T \cos\theta(t) dt & {\rm for} \ i\ne j.
\end{array}
\right.
\label{theta_5_23_22}
\end{eqnarray}
}

\begin{figure}[h!]
\begin{center}
\includegraphics[width=7cm]{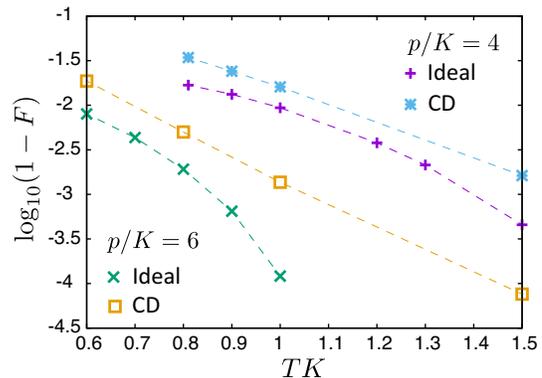}
\end{center}
\caption{
\textcolor{black}{$T$-dependence of the infidelity of the $R_{zz}$ gates based on the ideal beam-splitter-type coupling and the control with the CD term for $p/K=4$ and 6.
The initial state is $|\Psi_s\rangle$.
The value of $\theta_{\rm amp}$ was chosen so that  $(\varphi_{\bar{1}\bar{0}} - \varphi_{\bar{0}\bar{0}})/\pi=-0.5$.
The used parameters is $J/K=0.2$.
The dashed lines are guides to eyes.}
}
\label{fid_BS_com_7_3_22}
\end{figure}

\textcolor{black}{\section{Ideal tunable coupling}
\label{Beam splitter type}}
\textcolor{black}{
We consider the ideal tunable coupling with the form of $g(t)(a_1a_2^\dagger + a_1^\dagger a_2)$, which is called beam-splitter type~\cite{Gao2018}.
We compare the performance of our $R_{zz}$ gate with that of the $R_{zz}$ gate based on the ideal beam-splitter-type coupling.
For the control based on the ideal beam-splitter-type coupling, we set $g(t)=J\cos\theta(t)$ with $\theta(t) = \frac{\pi}{2} - \theta_{\rm amp} \pi [1-\cos(2\pi t/T)]$ for $0\le t \le T$, and fix $\theta_{1,2}$ to zero.
Figure~\ref{fid_BS_com_7_3_22} shows the infidelity of the $R_{zz}$ gates as a function of $T$.
The fidelity for the ideal beam-splitter-type coupling is higher than that of our scheme, although the difference is modest when the pump amplitude is small or $T$ is short.}

\textcolor{black}{
\section{Robustness}
\label{Robustness}}
\textcolor{black}{We examine the robustness of our scheme against the decoherence and errors in the CD term and in the resonance frequency of KPOs.}

\textcolor{black}{
\subsection{Decoherence}
\label{Decoherence}
}

\begin{figure}[h!]
\begin{center}
\includegraphics[width=7cm]{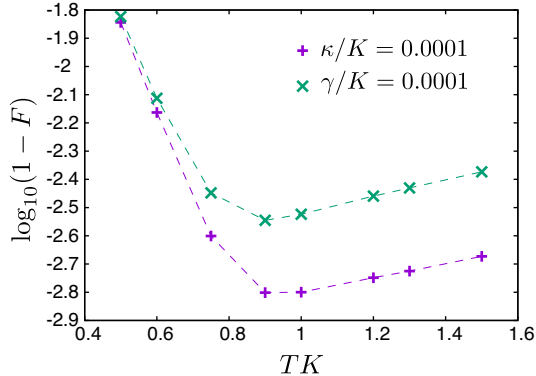}
\end{center}
\caption{
\textcolor{black}{Dependence of the infidelity on $T$ for the controls with the CD term 
for the cases of the pure decay with $\kappa_l=\kappa$ and pure dephasing with $\gamma_p^{(l)}=\gamma$.
The initial state is $|\Psi_s\rangle$.
The value of  $\theta_{\rm amp}$ was chosen so that  $(\varphi_{\bar{1}\bar{0}} - \varphi_{\bar{0}\bar{0}})/\pi=-0.5$
The used parameters are $p/K=7$ and $J/K=0.2$.
The dashed lines are guides to eyes.}
}
\label{fid_fin_com_purekappa_6_25_22}
\end{figure}

\textcolor{black}{
We consider the $R_{zz}$ gate with the CD term in the case that there is pure decay or pure dephasing.
Figure~\ref{fid_fin_com_purekappa_6_25_22} shows the dependence of the infidelity of the $R_{zz}$ gate on $T$.
It is seen that the $T$-dependence of the infidelity is quantitatively the same in the both cases.
The infidelity has a minimum with respect to $T$, where the suppression of nonadiabatic transitions balances with unwanted transitions due to the decoherence.
}

\begin{figure}[]
\begin{center}
\includegraphics[width=7cm]{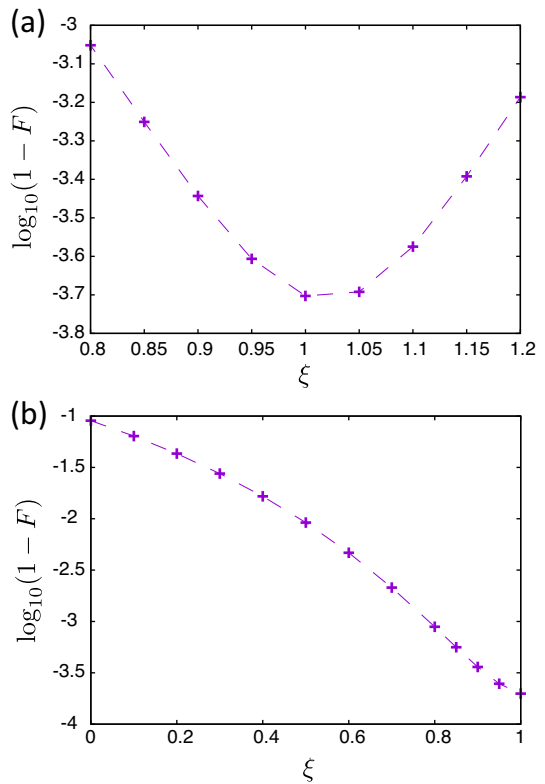}
\end{center}
\caption{
\textcolor{black}{$\xi$-dependence of the infidelity of the control with the CD term \textcolor{black}{for $0.8\le \xi\le 1.2$ (a) and for $0\le \xi\le 1$ (b)}.
The initial state is $|\Psi_s\rangle$.
The value of  $\theta_{\rm amp}$ was chosen so that  $(\varphi_{\bar{1}\bar{0}} - \varphi_{\bar{0}\bar{0}})/\pi=-0.5$.
The used parameters are $p/K=7$, $T/K=1$ and $J/K=0.2$.
The infidelity, ${\rm log}_{10}(1-F)$, of the control without the CD field is approximately $-1$.
}
}
\label{fid_error_5_22_22}
\end{figure}

\textcolor{black}{
\subsection{Imperfection of CD term}
\label{Imperfection of CD term}}
\textcolor{black}{The CD term may depart from the ideal one, $-\dot{\theta}_1(t)a_1^\dagger a_1$, due to imperfection of the control of the detuning.
In order to examine the robustness of our scheme against the error in the CD term, we assume that the imperfect CD term is represented as $-\xi\dot{\theta}_1(t)a_1^\dagger a_1$, where $\xi$ is the constant parameter characterizing the degree of the error.}
\textcolor{black}{Figure~\ref{fid_error_5_22_22} shows the dependence of the infidelity of the control with the CD term on $\xi$. 
We confirmed that the fidelity is significantly higher than that of the control without the CD term in the range of $\xi$ used.
Therefore, our scheme is robust against the error in the CD term.}
\textcolor{black}{As shown in Fig.~\ref{fid_error_5_22_22}(b), the infidelity monotonically decreases as $\xi$ increases for $0\le \xi\le 1$, where $\xi=0$ corresponds to the control without the CD term.
This result shows that our method can improve the control fidelity even with limited tunability of the detuning.}

\begin{figure}[]
\begin{center}
\includegraphics[width=7cm]{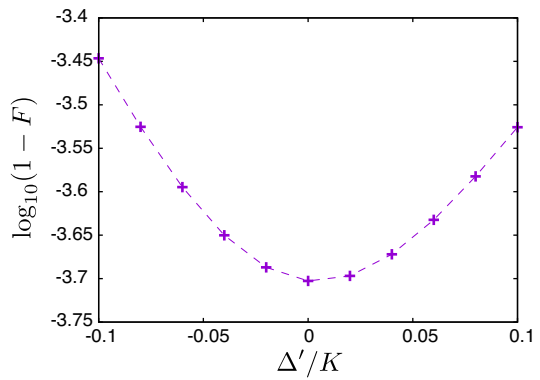}
\end{center}
\caption{
\textcolor{black}{Dependence of the infidelity on $\Delta'$ for the control with the CD term.
The initial state is $|\Psi_s\rangle$.
The value of  $\theta_{\rm amp}$ was chosen so that  $(\varphi_{\bar{1}\bar{0}} - \varphi_{\bar{0}\bar{0}})/\pi=-0.5$.
The used parameters are $p/K=7$, $T/K=1$ and $J/K=0.2$.
The dashed lines are guides to eyes.}
}
\label{fid_Delta_KPO1_6_30_22}
\end{figure}

\textcolor{black}{
In our scheme, the resonance frequencies of the KPOs are tuned to $\omega =  K + \omega_p/2$ when the coupling is off, so that the detuning is zero.
However, there might be an error in $\omega$.
We examine the robustness of our scheme against the error in $\omega$.
In order to describe the error of $\omega$, we introduce an additional constant detuning, $\Delta'$, of KPO~2. 
Figure~\ref{fid_Delta_KPO1_6_30_22} shows the dependence of the infidelity on $\Delta'$.
It is seen that the fidelity of the control is not sensitive to the discrepancy of the resonance frequency of the KPO 2 in the parameter range studied.
}

\textcolor{black}{
\section{Loading into ground state}
\label{Loading into ground state}
}
\textcolor{black}{
A KPO can be loaded to its ground state, which is the cat state, from the vacuum state by adiabatically ramping the pump amplitude~\cite{Cochrane1999,Goto2016}.
The control fidelity is degraded due to nonadiabatic transitions, and the degradation is enhanced when the duration of the control becomes short.
Time dependent detuning can suppress nonadiabatic transitions, thus it can be used to shorten the duration of the loading process~\cite{Goto_patent,Masuda2020}.
We examine the efficiency of these adiabatic loading protocols in our two-KPO system.
We assume that the coupling is set to be off, that is, $\theta=\pi/2$ throughout the initialization, and the decoherence is negligible (the effect of the decoherence is studied for a KPO e.g. in Ref.~[\citenum{Masuda2020}]).
The pump amplitude is monotonically increased from zero to $p_{\rm max}$ for $0\le t \le T$ as
$p_l(t)=p_{\rm max}[1 - \cos(\pi t/T) ] / 2$ for $l=1,2$. We set $p_l/K=4$ and $J/K=0.2$.
}

\begin{figure}
\begin{center}
\includegraphics[width=7.5cm]{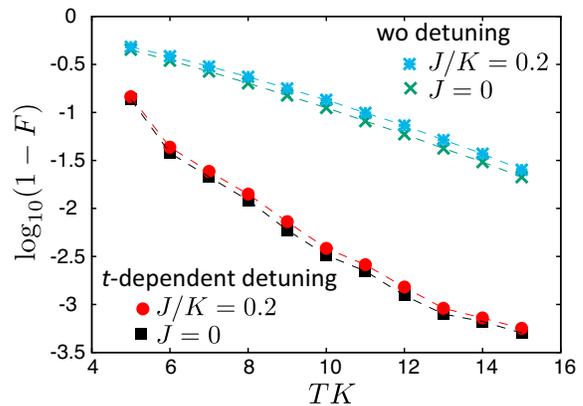}
\end{center}
\caption{
\textcolor{black}{
Dependence of the infidelity of the adiabatic loading protocols on the loading time $T$ for $p/K=4$. 
The light blue asterisks and green crosses are for the control without detuning,
while the others are for the control with the time dependent detuning.
The light blue asterisks and the red filled circles are for $J/K=0.2$, while the others are for $J=0$.
The dashed lines are guides to eyes.
}
}
\label{fid_p4_com_6_25_22}
\end{figure}

\textcolor{black}{
We first examine the loading protocol with the detuning fixed to zero.
The control fidelity is defined by the squared amplitude of the overlap between the state at $t=T$ and the  ground state of $H_{\rm tot}(T)$ in Eq.~(\ref{H_2KPO_11_25_21}) with $\dot\theta=0$.
For comparison, the fidelity of the control without the always-on linear coupling ($J=0$) is also presented, where the fidelity is defined with the ground state for $J=0$.
The squared amplitude of the overlap between the two ground states corresponding to $J/K=0$ and $0.2$ is 0.994. 
As seen in Fig.~\ref{fid_p4_com_6_25_22}, the fidelities of the controls for $J/K=0$ and $0.2$ are approximately the same for the parameters used. 
}

\textcolor{black}{
Next, we consider the loading protocol with the time dependent detuning.
The role of the detuning is to open the gap between energy levels of each KPO to mitigate unwanted nonadiabatic transitions, and thus the role is different from that of the detuning used in the main text. 
The time dependence of the detuning is given by $\Delta_l(t)=\Delta_{\rm max}[1 + \cos(\pi t/T) ] / 2$.
The detuning monotonically decreases from $\Delta_{\rm max}$ to zero for $0\le t\le T$. 
We set $\Delta_{\rm max}/K=3$ in this paper.
Figure~\ref{fid_p4_com_6_25_22} shows that fidelity of the control for $J/K=0.2$ is greatly improved
as well as that for  $J=0$ compared to the fidelity of the control without detuning.
}


\end{document}